\documentclass{aa}  

\usepackage{graphicx}
\usepackage{amssymb,amsmath}
\usepackage{latexsym}
\usepackage{hyperref}								
\usepackage{lmodern}
\usepackage{mathtools}
\usepackage{epsfig}
\usepackage{color}
\usepackage{cancel}
\usepackage{natbib}
\usepackage{needspace}
\usepackage[varg]{txfonts}
\usepackage[utf8]{inputenc}

\makeatletter
\renewcommand*\aa@pageof{, page \thepage{} of \pageref*{LastPage}}
\makeatother

\begin{document}

\title{Origin of asteroid (469219) Kamo`oalewa: the main asteroid belt or the Giordano Bruno crater on the Moon?}

   \author{Marco Fenucci\inst{1,2}
          \and
          Bojan Novakovi\'{c}\inst{3,4}
          \and
          Mikael Granvik\inst{5,6}
          \and
          Pengfei Zhang\inst{7}
          }
          
   \authorrunning{M. Fenucci et al.}
    \institute{ESA ESRIN / PDO / NEO Coordination Centre, Largo Galileo Galilei, 1, 00044 Frascati (RM), Italy \\ \email{marco.fenucci@ext.esa.int}
    \and
    Deimos Italia s.r.l., Via Alcide De Gasperi, 24, 28060 San Pietro Mosezzo (NO), Italy
    \and
    Department of Astronomy, Faculty of Mathematics, University of Belgrade, Studentski trg 16, Belgrade, 11000, Serbia
    \and
    Centro de Estudios de Física del Cosmos de Aragón (CEFCA), Plaza San Juan 1, 44001 Teruel, Spain
    \and
    Department of Physics, University of Helsinki, P.O. Box 64, 00014, Finland
    \and
    Asteroid Engineering Laboratory, Luleå University of Technology, Box 848, 981 28 Kiruna, Sweden
    \and
    Center for Lunar and Planetary Sciences, Institute of Geochemistry, Chinese Academy of Sciences, Guiyang (Guizhou), PR China
    }

   \date{Received --- / Accepted ---}

  \abstract
  % context heading (optional)
     {Asteroid (469219) Kamo`oalewa is the target of the Tianwen-2 sample-return mission by the China National Space Administration. Because of its orbit and its peculiar spectral properties, it was proposed that Kamo`oalewa originated from the Moon as impact ejecta, possibly from the Giordano Bruno crater.}
  % aims heading (mandatory)
     {We aim at estimating the relative contribution of Kamo`oalewa-like objects originating from the general near-Earth asteroid (NEA) population which originated in the main asteroid belt, and compare it with the relative contribution of Giordano Bruno ejecta.}
  % methods heading (mandatory)
     {We first estimate the average fraction of quasi-satellite orbits in the Earth co-orbital space at any given time by using numerical simulations. By using recently developed NEA population models, we extract the expected number of Earth co-orbitals of the same size of Kamo`oalewa, and then we get an estimate of the average number of Kamo`oalewa-like objects using the fraction computed before. Similarly, we obtain an estimate for the number of Kamo‘oalewa-like objects that may originate as ejecta from the Giordano Bruno impact. We also performed simulations for the Catalina Sky Survey, Pan-STARRS, and Vera Rubin Observatory, to estimate their efficiency in the detection of Kamo`oalewa-like objects. }
  % results heading (mandatory)
     {Numerical simulations showed that 1.39\% of the orbits in the Earth co-orbital space are quasi-satellite, on average. When combined with the expected number of Earth co-orbitals in the same size range of Kamo`oalewa from NEA population models, we found that the main belt accounts for $1.23 \pm 0.13$ Kamo`oalewa-like objects on average. The expected number of Kamo`oalewa-like objects originated as Giordano Bruno ejecta is 0.042, which is more than order of magnitude smaller. On the other hand, we found a discovery efficiency  of Earth quasi-satellites between 95\% and 70\% for absolute magnitude between 22 and 25 for the Pan-STARRS survey, and population models show that this is in agreement with the known population of Earth quasi-satellites. The Vera Rubin Observatory should reach a discovery efficiency of 92\% down to absolute magnitude 25.}
  % conclusions heading (optional)
     {Quantitative estimates show that population models of NEAs based on the migration of objects from the main belt are capable to account for Kamo`oalewa-like objects. This relative contribution supports the hypothesis that (469219) Kamo`oalewa originated from the main belt, which will be further investigated by future observations and in-situ exploration of the Tianwen-2 spacecraft. } 
     
     \keywords{minor planets, asteroids: individual: (469219) Kamo`oalewa - methods: statistical}

%%%%%%%%%%%%%%%%%%%%%%%%%%%%%%%%%%%%%%%%%%%%%%%%%%%%%%%%%%%%%%%%%%%%%%%%%%%%%%%%%%%%%%%%%%%

\maketitle

\section{Introduction}
\label{s:intro}
Near-Earth asteroid (NEA) (469219) Kamo`oalewa is the target of the Tianwen-2 sample-return mission \citep{zhang-etal_2021, zhang-etal_2025natas} of the China National Space Administration (CNSA), which was successfully launched on 29 May 2025. Kamo`oalewa has peculiar characteristics in both its orbit and known physical properties. This object is in a 1:1 mean-motion resonance with the Earth, and it is currently in a quasi-satellite configuration \citep{delafuente-delafuente_2016}. On a timescale of 10 kyr it repeatedly switches between quasi-satellite and horseshoe configuration \citep{delafuente-delafuente_2016, fenucci-etal_2022}, and its dynamical stability properties makes it a companion of Earth for at lest the next 0.5 Myr \citep{fenucci-novakovic_2021}. 
The available photometric measurements indicate an absolute magnitude of $H=24$, meaning that its diameter can be roughly estimated between 30 and 100 m in diameter, and its rotational period is only about 28 minutes \citep{sharkey-etal_2021}. Because of the long observational arc it was possible to determine the Yarkovsky effect \citep{liu-etal_2022, fenucci-etal_2025} with a good signal-to-noise ratio. This permitted to evaluate also the thermal inertia with the ASTERIA method \citep{novakovic-etal_2024}, indicating a value of $150^{+90}_{-45}$ J m$^{-2}$ K$^{-1}$ s$^{-1/2}$  \citep{fenucci-etal_2025}.
Initial spectral data in the visible and near-infrared wavelengths suggested that Kamo`oalewa is either an S-type or L-type asteroid \citep{reddy-etal_2017}. Additional measurements in the zJHK colors revealed a high red slope \citep{sharkey-etal_2021} typical of materials that suffered prolonged exposure to space weathering effects \citep{pieters-etal_1993}. 

The peculiar red slope also resembles that of lunar regolith, which, together with its small size, yields the hypothesis that  Kamo‘oalewa originated from the Moon as ejecta from an ancient asteroid impact. Numerical simulations showed that there are rare pathways bringing lunar ejecta into the Earth co-orbital space \citep{castro-etal_2023}, which can later transition to a quasi-satellite configuration. By studying the age of the craters on the Moon and their sizes, the Giordano Bruno crater was proposed as the only possible crater from which Kamo`oalewa could originate \citep{jiao-etal_2024}. Statistics presented in \citeauthor{jiao-etal_2024} showed that the impact which created the Giordano Bruno crater could produce 0.3$-$1 Earth co-orbitals on average. These fragments later need to transition into a quasi-satellite state in order to match the dynamical properties of Kamo`oalewa.  
\citet{jiao-etal_2024} also compared the available spectral data of Kamo`oalewa to the Apollo 14 and the Luna 24 mission samples, as well as a representative Lunar meteorite. In particular, the Luna 24 sample has been suggested to contain ejecta material from the Giordano Bruno crater \citep{basilevsky-head_2012}. In the comparison made using the mixture of common factor analyses, the spectrum of Kamo`oalewa appears to be similar to those of the Apollo 14 samples and the Lunar meteorite. On the other hand, the Luna 24 sample is the farthest one from Kamo`oalewa, which seems in contradiction with the fact that this sample is supposed to contain Giordano Bruno ejecta. 

On the other hand, the spectrum of the Giordano Bruno crater was extensively studied by \citep{ogawa-etal_2011} by using data from the Japanese Selenological and Engineering Explorer (SELENE) lunar orbiter. The authors presented the reflectance spectrum between 0.5 and 2 $\mu$m wavelengths of several locations across the crater, including the field of ejecta right outside the crater, the rim, the wall, and the floor of the crater itself. All the examined locations show two prominent absorptions bands at 1 $\mu$m and at 1.3 $\mu$m, with significant variations between these two wavelengths depending on the location. Moreover, the steep red slope is not present in any of these data analysed, raising further questions on the origin of Kamo`oalewa.

Laboratory analyses showed that laser irradiated LL-chondrite powder (size smaller than 45 $\mu$m), which simulates the effect of space weathering, can reach a steep spectral slope \citep{zhang-etal_2024}, suggesting that Kamo`oalewa could be an asteroid originated in the main belt which suffered long exposure to space weathering. The same study suggested that LL chondrites, which are thought to originate from the Flora family \citep{vernazza-etal_2007}, are meteorite analogues for Kamo`oalewa. However, the contribution of the main belt to small NEAs reaching a quasi-satellite state with the Earth has not yet been estimated. 

These considerations underscore the need for cautious interpretation of spectral data, and that other possibilities should be considered $-$ together with the statistics coming from the possible dynamical evolution. The upcoming observations of Kamo`oalewa by the James Webb Space Telescope (JWST) with the Near Infrared Spectrograph (NIRSpec) planned in the first quarter of 2026 \citep{sharkey-etal_2025} will provide additional high resolution spectral data, together with better constraints on its size and albedo. JWST has recently demonstrated exceptional capabilities in asteroid studies \citep{burdanov-etal_2025, rivkin-etal_2025, rivkin-etal_2025b}, and this new dataset is expected to better establish the comparisons with lunar samples ahead of the arrival of the Tianwen-2 spacecraft.

In this work we estimate the number of Kamo`oalewa-like objects $-$ i.e. Earth quasi-satellites with a size similar to that of Kamo`oalewa $-$ originating from the main belt. The relative contribution of ejecta from the Giordano Bruno crater is also evaluated for comparison. We also perform survey simulations to give quantitative estimates of the expected completeness of the population of Kamo`oalewa-like objects.

\section{Methods}
\label{s:methods}

\subsection{Average fraction of Earth quasi-satellites}
We define the Earth co-orbital space as orbits with osculating semi-major axis $a$, eccentricity $e$, and inclination $i$ such that
\begin{equation}
    0.994 \text{ au} < a < 1.006 \text{ au}, \ 0 < e < 0.2, \ 0 \text{ deg} < i < 20 \text{ deg}.
    \label{eq:coorbital_space}
\end{equation}
On the other hand, we do not put constraints on the longitude of the node $\Omega$, the argument of the pericenter $\omega$, and the mean anomaly $M$. 
The resonant dynamics is then identified by the time evolution of the angle  $\sigma = \lambda - \lambda_E = (\Omega + \omega + M) - (\Omega_E + \omega_E + M_E)$, where the subscript $E$ denotes the corresponding orbital element of the Earth.

Note that, while orbits with larger eccentricity and inclination could still be Earth co-orbitals, the constraints of Eq.~\eqref{eq:coorbital_space} are consistent with the secular evolution of Kamo`oalewa shown in \citet{delafuente-delafuente_2016, fenucci-etal_2022}. In fact, over 10 kyr of dynamical evolution, which is a timespan after which the orbit of Kamo`oalewa becomes stochastic, the semi-major axis of Kamo`oalewa oscillates between the 0.995 au and 1.005 au, while the eccentricity does not exceed 0.14 and the inclination does not grow larger than 10 deg.  

To estimate the fraction of Earth co-orbitals which are in a quasi-satellite state we proceed with a Monte Carlo approach. We sample the Earth co-orbital space with a large number $N$ of initial conditions and numerically propagate them for a time $T$. All the initial orbital elements are assumed to be uniformly distributed within their domain of definition, since the whole phase space has to be explored. For each initial condition, we determine the time $T_{QS}$ spent in a quasi-satellite state, and we then compute the average fraction spent in a quasi-satellite state as
\begin{equation}
    F = \frac{1}{N} \sum_{k=1}^N \frac{T_{QS}^{(k)}}{T}.
    \label{eq:avg-frac}
\end{equation}
The number $F$ provides the average fraction of quasi-satellites within the Earth co-orbital space at any given time. An averaged approach is needed because Earth co-orbitals may switch their state multiple times, as it is also seen in the dynamics of Kamo`oalewa.

\subsection{Numerical simulations and quasi-satellite identification}
\label{s:simulations}
We fixed the initial epoch at 60800 MJD, and initial conditions for all the planets from Mercury to Neptune were obtained by using the JPL Ephemeris DE441 \citep{park-etal_2021}. 
Note that the resonant averaged motion is determined only by the initial $a$ and $\sigma$ \citep{fenucci-etal_2022, pan-gallardo_2025} and, since there are no constraints on $\omega, \Omega$ and $M$, all the possible values of $\sigma$ are explored by the Monte Carlo approach. Thus, fixing a common starting epoch for the numerical simulations is fully justified by the properties of the resonant averaged dynamics, even if we are not using an averaged dynamical model for the simulations.  
Numerical integrations were performed with a hybrid symplectic integrator \citep{chambers_1999}, which is able to solve close encounters with planets while maintaining the symplectic characteristics of the integration scheme. This integration method is included in the \texttt{mercury} package \citep{chambers-migliorini_1997}. While the Yarkovsky effect could be added to the dynamical model of \texttt{mercury} \citep{fenucci-novakovic_2022}, it has been shown that it is not relevant in the short term dynamical evolution of Kamo`oalewa \citep{fenucci-novakovic_2021}, hence we do not include it in the model. We used a timestep of 0.5 days for the numerical integrations, and the dynamical state was recorder with a step of 1 year, which is enough to determine the resonance type of the orbit. 

\begin{figure}
    \centering
    \includegraphics[width=0.48\textwidth]{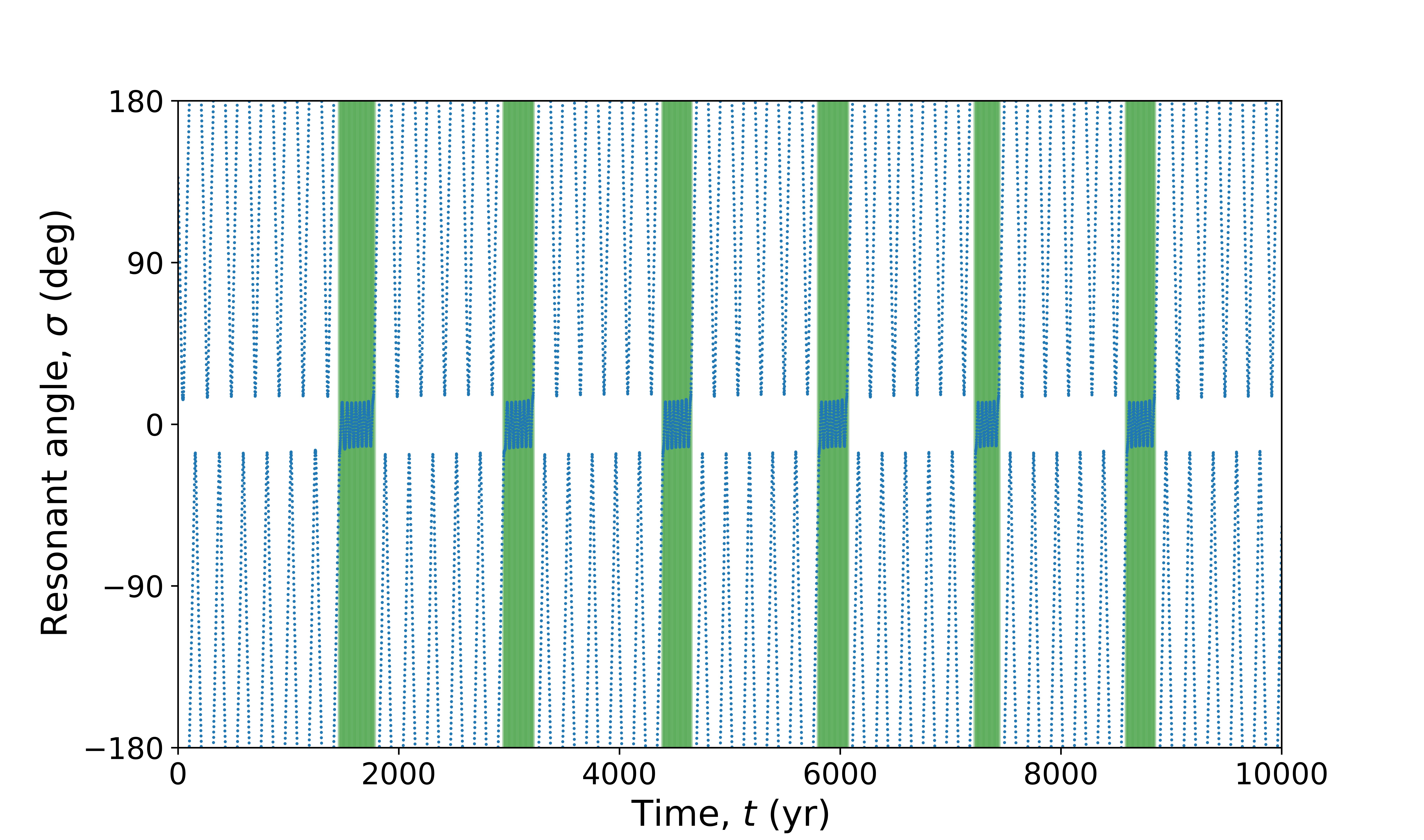}
    \caption{Time evolution of the resonant angle $\sigma$ (blue dots) of an example orbit which switches between horseshoe and quasi-satellite states. The green shaded areas are the quasi-satellite states identified with the running-window procedure. }
    \label{fig:qs_running_window}
\end{figure}

The time spent in a quasi-satellite state $T_{QS}$ is determined by a running-windows approach. First, we compute the time evolution of the resonant argument $\sigma$. To identify a quasi-satellite state we check the following conditions on $\sigma$ and the semi-major axis $a$: 1) $-50 \text{ deg} < \sigma < 50 \text{ deg}$; 2) $a$ must attain values larger and smaller than 1 au at different epochs; 3) $\sigma$ must have at least 2 local minima and 2 local maxima. Using a fixed time window with length $T_W$, all the intervals $[t_k, t_k + T_W]$ are checked for quasi-satellite states, where $t_k$ are the output times from the numerical integrations. To catch different libration periods, we scan the whole integration timespan with running-windows of length $T_W = 10, 20, 50, 100, 150, 200, 250, 300, 350$ years and $T_W = 400, 500, \dots, 1000$ years. The time $T_{QS}$ is then obtained as the sum of the quasi-satellite instances found. Figure~\ref{fig:qs_running_window} shows the time evolution of the resonant angle $\sigma$ for an example orbit, with the green-shaded areas indicating the quasi-satellite states identified by the running-windows procedure.

\subsection{Expected number of Kamo`oalewa-like objects}
\label{ss:N_Kamooalewa}
The expected number $N_{\text{coorb}}$ of Earth co-orbitals in the same size range of Kamo`oalewa can be obtained by NEA population models, such as those by \citet{granvik-etal_2018} or \citet{nesvorny-etal_2024, nesvorny-etal_2024b}. These models provide a size-frequency distribution (SFD) of NEAs, as well as the expected distribution of their orbital elements $(a,e,i)$, thus allowing filtering with the constraints of Eq.~\eqref{eq:coorbital_space}. We can also assume a Poisson statistics on $N_{\text{coorb}}$, therefore its standard deviation is $\sigma(N_{\text{coorb}}) = \sqrt{N_{\text{coorb}}}$. Note that these NEA population models are based on the migration of asteroids from the main belt to the NEA region by the combined action of resonances and the Yarkovsky effect, thus all the predicted NEAs are assumed to originate in the main asteroid belt \citep{granvik-etal_2017}. 
The average number of Kamo`oalewa-like objects\footnote{With the term Kamo`oalewa-like object we mean an NEA which is currently a quasi-satellite of the Earth and with the same size of Kamo`oalewa.} in the steady-state NEA population is then obtained as 
\begin{equation}
    N_{\text{K/NEA}} = F \times N_{\text{coorb}}. 
    \label{eq:NKNEA}
\end{equation}

Similarly, we can evaluate the relative contribution of Kamo`oalewa-like objects originated as ejecta from the Giordano Bruno crater. \citet{jiao-etal_2024} showed that the impact which created the Giordano Bruno crater would produce about 300 fragments of the same size as Kamo`oalewa, and that 0.0001$-$1\% of them have a chance to become Earth co-orbitals at 3$\sigma$ confidence level. Thus, the most optimistic estimate for the number of Earth co-orbitals from Giordano Bruno is $N_{\text{GB}} = 3$. Therefore the average number of Kamo`oalewa-like objects from the Giordano Bruno crater is $N_{\text{K/GB}} = F \times N_{\text{GB}}$.

\subsection{Survey discovery efficiency of Kamo`oalewa-like objects}
\label{ss:surveyeffmethod}
We perform survey simulations to understand the efficiency in discovering Kamo`oalewa-like objects. We focus on the two most productive currently operational surveys, i.e. Pan-STARRS \citep[MPC code F51,][]{denneau-etal_2013} and the Catalina Sky Survey \citep[CSS, MPC code G96,][]{fuls-etal_2023}. We also simulate the efficiency for the upcoming Vera Rubin Telescope \citep{ivezic-etal_2019}. 
Given a set of $N$ asteroids which are currently in a quasi-satellite orbit with the Earth, with orbital elements fulfilling Eq.~\eqref{eq:coorbital_space} and absolute magnitude $22<H<25$, we simulate the observational ephemerides from a given survey using the Aegis software \citep{fenucci-etal_2024b} by the European Space Agency (ESA). Ephemerides for G96 are simulated from 1 January 2005 to 1 October 2025, while for F51 from 1 January 2011 to 1 October 2025, thus covering the whole operational period of these two surveys. For Vera Rubin, we compute ephemerides from 1 January 2026 to 1 January 2036, which roughly corresponds to its expected operational lifetime. 

To simulate the survey strategy to scan the observable sky and to take into account full moon nights and realistic weathering effects, we use the available telescope pointing data. Historic pointing data for G96 and F51 are available at the Minor Planet Center\footnote{\url{https://www.minorplanetcenter.net/iau/info/PointingData.html}} (MPC), although those of F51 are available only until 2022. These are real pointings, and they already take into account real observing conditions experienced in the last decades of operations. Simulated pointing data for the next 10 years are also available for the Vera Rubin Observatory, which are freely available\footnote{\url{https://survey-strategy.lsst.io/baseline/}}. For the simulations done in this work, we used the latest database of pointings called \texttt{baseline\_v5.0.0\_10yrs}.
We also assume a limiting magnitude of 21.5 for G96, 22.0 for F51, and 24.0 for Vera Rubin, and assume that every object moving faster than 10 deg day$^{-1}$ are not detected by automated pipelines \citep{jedicke-etal_2025}. We also simulate the actual visual magnitude of an object observed by a telescope as $V + \Delta V$, where $V$ is the visual magnitude computed from the ephemerides with the $HG$ model using $G=0.15$ \citep{bowell-etal_1989}, and $\Delta V$ is a factor due to trailing loss. The $\Delta V$ term depends on the plane-of-sky velocity of the object and from the characteristics of the telescope, and we compute it as described in \citet{nesvorny-etal_2024}.

Given the ephemerides of an object at a certain epoch, we consider it detectable by the survey if: 1) the simulated visual magnitude $V+\Delta V$ is smaller than the limiting magnitude; 2) the corresponding Right Ascension (RA) and Declination (Dec) coordinates fall within one of the pointings of the survey of that same epoch; 3) the plane-of-sky velocity is smaller than 10 deg day$^{-1}$. Running the simulations of a sample of Earth quasi-satellites over the operational lifetime of the survey and computing the cumulative distribution of the detectable objects, gives the efficiency in discovering Kamo`oalewa-like objects. 

\section{Results}

\subsection{Expected number of Kamo`oalewa-like objects}
\label{s:results}
\begin{figure}
    \centering
    \includegraphics[width=0.48\textwidth]{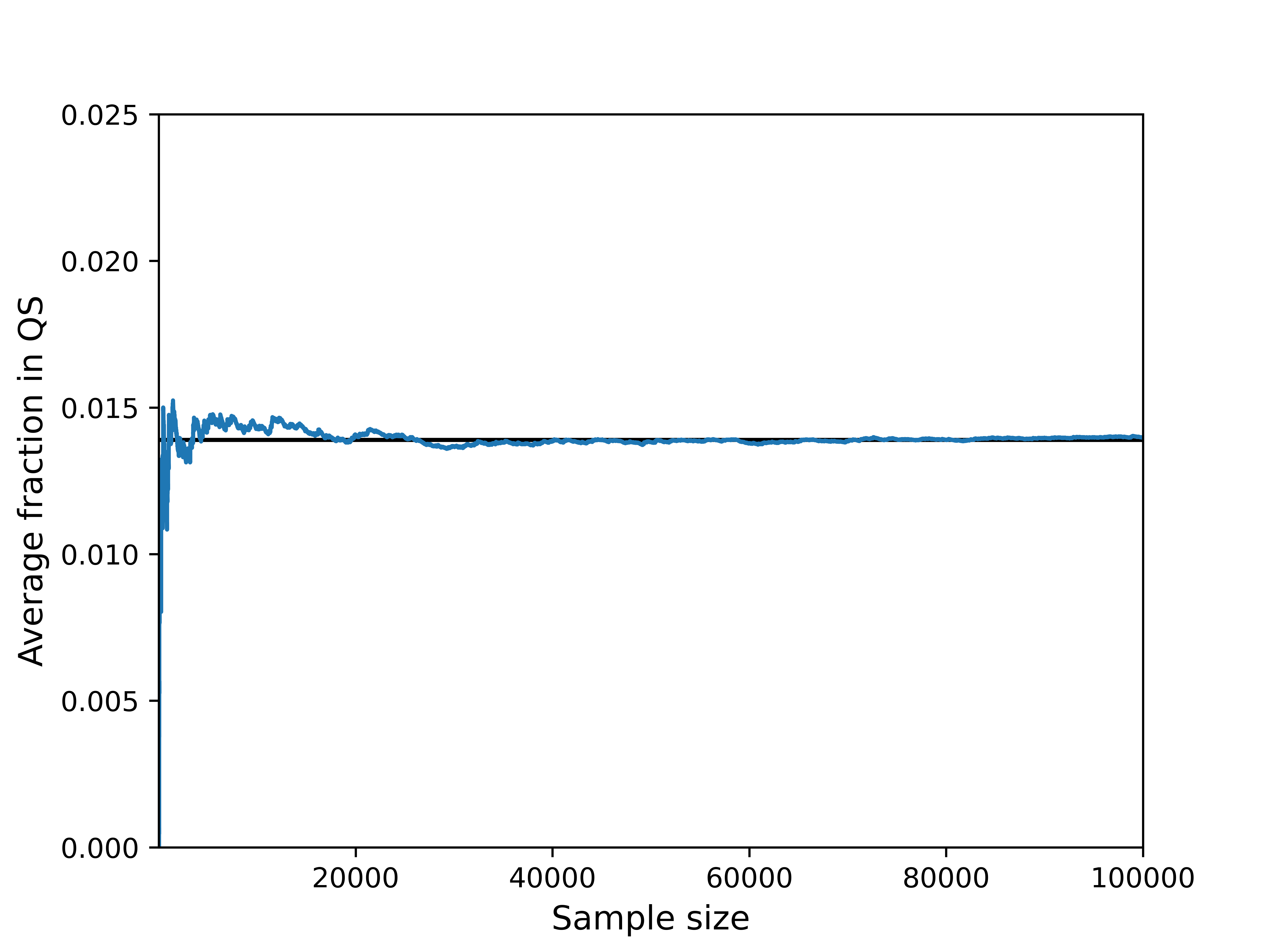}
    \caption{The running percentage of the average fraction of quasi-satellites, computed on an increasingly large sample of orbits.}
    \label{fig:percentage_stability}
\end{figure}

We sampled the Earth co-orbital space defined in Eq.~\eqref{eq:coorbital_space} with $N = 100~000$ randomly generated orbits, and their dynamics was propagated for 10 kyr. To verify that the chosen sample is large enough, we computed the average fraction of quasi-satellites on an increasingly larger sample. Figure~\ref{fig:percentage_stability} shows the results of these computations.
When the sample is smaller than about 20~000 orbits, the average fraction of quasi-satellites undergoes some fluctuations, while it stabilises for larger samples and approaches a limit  when $N > 80~000$, obtaining $F = 0.01396$ for the complete sample. This proves that our estimate is not dependent on the size of a chosen sample, provided it is large enough. The value obtained means that, on average, only 1.396\% of the objects in the Earth's co-orbital space (Eq.~\eqref{eq:coorbital_space}) are quasi-satellite at any given time. 
To verify that $F$ also does not depend on the total integration time, we extended all the integrations up to 50 kyr. 
The final value of $F$ is 0.01224, a change of about 10\% and only slightly smaller than that obtained by the 10 kyr integrations, thus indicating that the result does not strongly depend on the total integration timespan. 

As outlined in Sec.~\ref{ss:N_Kamooalewa}, the number of Kamo`oalewa-sized Earth co-orbitals originated as ejecta from the Giordano Bruno crater has been estimated to be $N_{\text{GB}} = 3$ in the best case scenario, thus the corresponding average number of Kamo`oalewa-like objects is $N_{\text{K/GB}} = 0.042$.  

\begin{figure}
    \centering
    \includegraphics[width=0.48\textwidth]{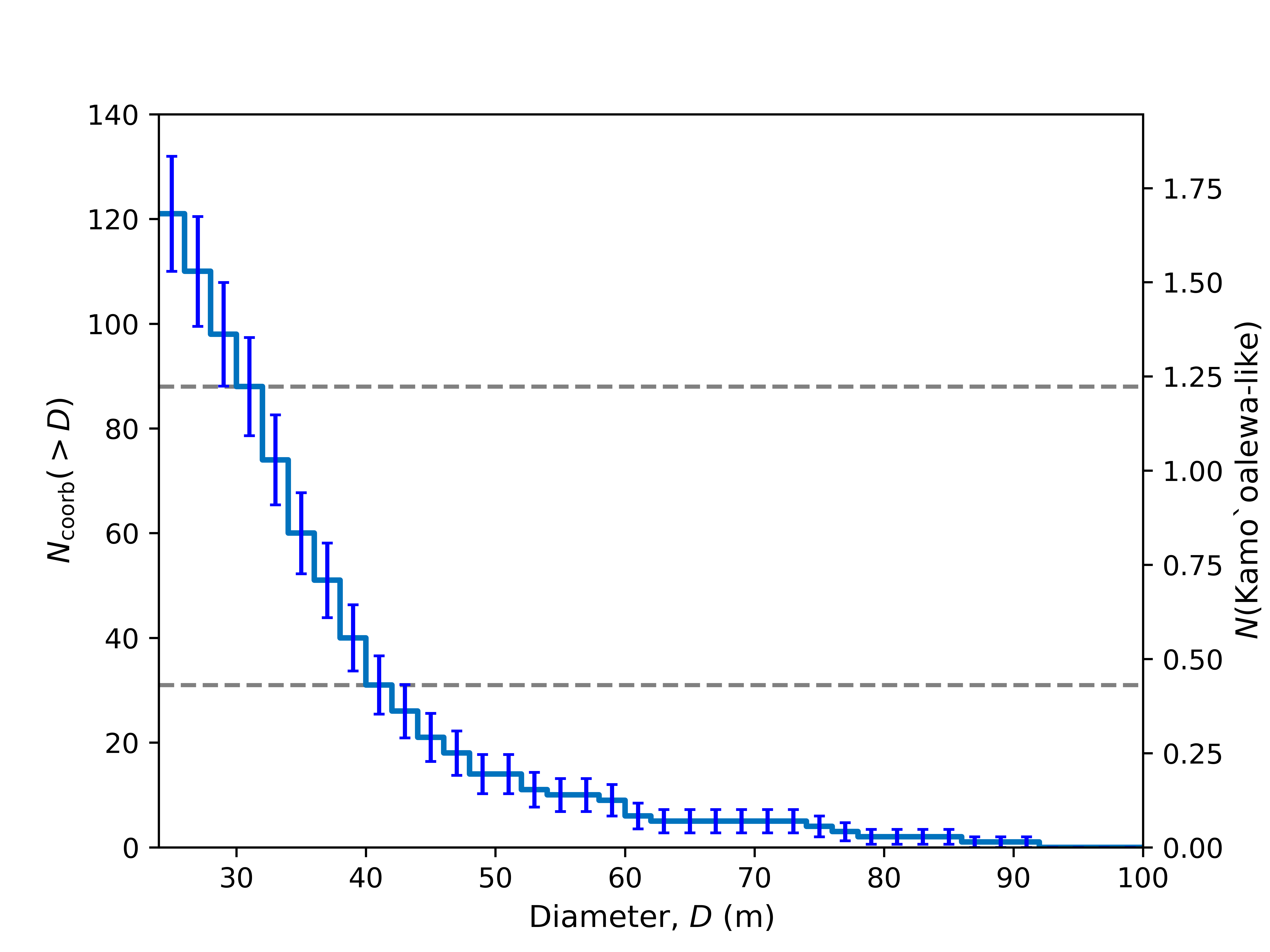}
    \caption{Cumulative size frequency distribution of NEAs between 30 m and 100 m in the Earth co-orbital region, extracted from the NEOMOD3 NEA population model. Error bars are computed assuming a Poisson statistics. The right y axis scales to the estimated number of Kamo`oalewa-like objects. Grey horizontal dashed lines highlight the cases with diameter of 30 and 40 m.}
    \label{fig:NEOMOD3_distribution}
\end{figure}

As described above, we can evaluate the expected number of Kamo`oalewa-like objects from NEA population models. Since we are interested in objects similar to Kamo`oalewa also in size\footnote{Recall that the absolute magnitude of Kamo`oalewa is $H=24$.}, we filter on absolute magnitude values $H$ between 22 and 25. With these constraints, the model by \citet{granvik-etal_2018} estimates $N_{\text{coorb}} = 88 \pm 9.3$, while the NEOMOD2 model by \citet{nesvorny-etal_2024} gives $N_{\text{coorb}} = 73 \pm 8.5$. The diameter of Kamo`oalewa is not precisely known, but a-priori models based on photometry and population models indicate a size between 30 m and 100 m \citep{fenucci-etal_2025}. The NEOMOD3 model by \citet{nesvorny-etal_2024b} provides also the size distribution of the NEA population, and the constraint 30 m $<D<$ 100 m can also be used to extrapolate the number of NEAs in the Earth co-orbital region. Figure~\ref{fig:NEOMOD3_distribution} shows the cumulative SFD of Earth co-orbitals between 25 m and 100 m, extracted from a sample NEA population generated with NEOMOD3. Using the size constrain of Kamo`oalewa, we get $N_{\text{coorb}} = 88 \pm 9.3$, consistent with the model by \citet{granvik-etal_2018}. 

By using the estimate outlined in Sec.~\ref{s:methods}, we obtain an average number $N_{\text{K/NEA}} = 1.23 \pm 0.13 $ of Kamo`oalewa-like objects with $22 < H < 25$ and with 30 m $<D<$ 100 m, using respectively the \citeauthor{granvik-etal_2018} and the NEOMOD3 models. We obtain instead $N_{\text{K/NEA}} = 1.02 \pm 0.12$ Kamo`oalewa-like objects with $22 < H < 25$ by assuming the NEOMOD2 model. The right axis of Fig.~\ref{fig:NEOMOD3_distribution} shows also how the number of Kamo`oalewa-like objects varies for different values of the diameter using the NEOMOD3 model as input distribution. 

\subsection{Survey discovery efficiency of Kamo`oalewa-like objects}
\label{ss:completion_results}
To obtain a sample of objects which are currently quasi-satellites of the Earth, we randomly generated orbits with initial Keplerian elements fulfilling Eq.~\eqref{eq:coorbital_space} at epoch 60800 MJD, and with an initial resonant angle $\sigma$ between $-50$ and 50 deg. The orbits were propagated for 50 years both in the past and in the future, and only those which were quasi-satellites of the Earth for this whole timespan were kept. With this procedure, we generated a set of $N=5000$ Earth quasi-satellites, which were assigned a random absolute magnitude $H$ between 22 and 25. We then simulated the survey efficiency with the settings described in Sec.~\ref{ss:surveyeffmethod}.

\begin{figure*}
    \centering
    \includegraphics[width=0.98\linewidth]{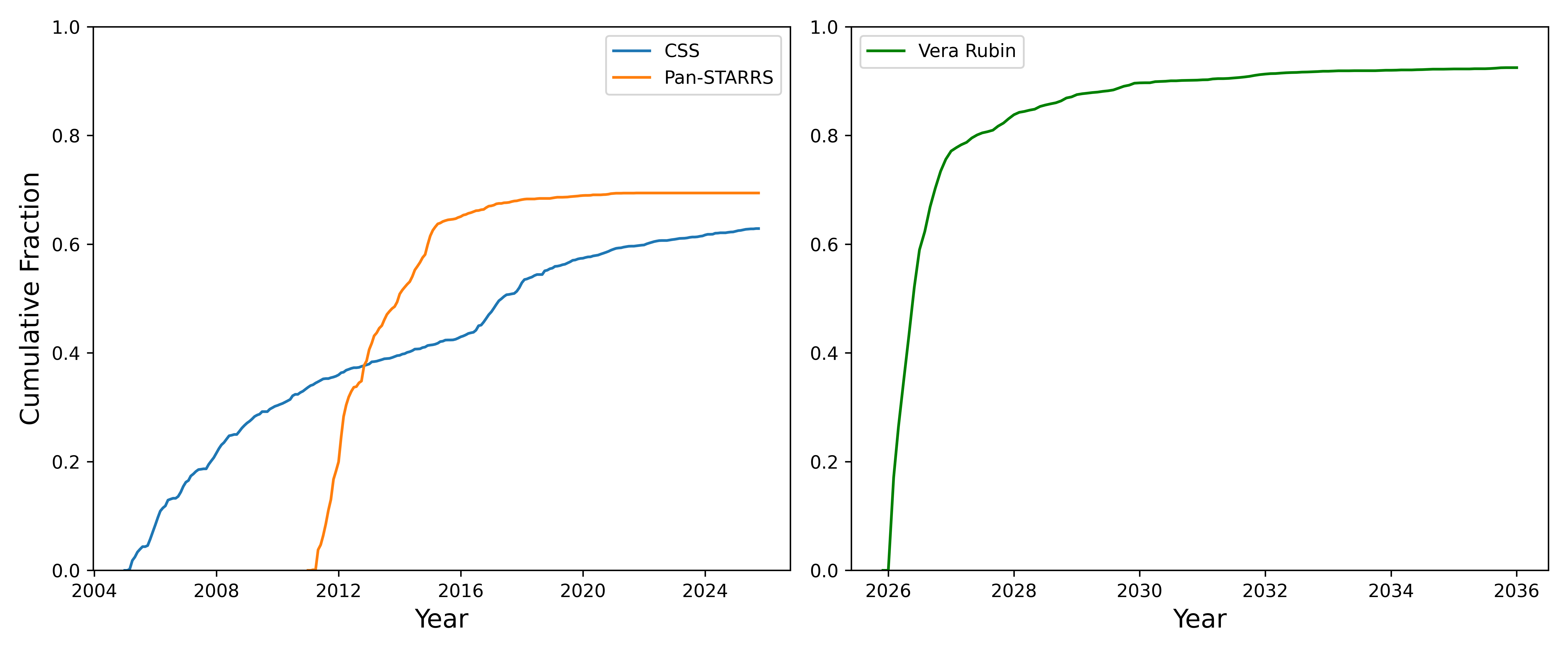}
    \caption{Cumulative fraction of detected Kamo`oalewa-like quasi-satellites as a function of the year. The left panel shows the results obtained for the CSS survey (blue curve) and for the PanSTARRS survey (orange curve). The right panel shows the expected efficiency of the Vera Rubin Telescope during its 10 years operations period.}
    \label{fig:survey_total_detections}
\end{figure*}

The left panel of Fig.~\ref{fig:survey_total_detections} shows the cumulative fraction of discovered Kamo`oalewa-like objects by CSS and Pan-STARRS, as a function of the year. CSS shows an almost linear trend in the number of discoveries, ultimately arriving to about 62\% after almost 20 years of operations. The biggest limitation of CSS in discovering Kamo`oalewa-like objects is in the limiting visual magnitude, which is often not sensible enough to spot faint Earth quasi-satellites. On the other hand, the number of discoveries by Pan-STARRS is very steep in the first 4 years of operations, and the distribution becomes almost shallow after 8 years of operations. The final fraction of discovered Kamo`oalewa-like objects is about 70\%.
Results for the Vera Rubin observatory are reported in the right panel of Fig.~\ref{fig:survey_total_detections}. The behaviour is similar to that of Pan-STARRS: a large fraction of Kamo`oalewa-like objects are discovered within the first 4 years of operations, with few new discoveries in the years after. Our results indicates that Vera Rubin is expected to reach a completeness limit of 90\% by 2030, and 92\% by 2036. Thus, any undiscovered Kamo`oalewa-like object would be almost certainly discovered by the Vera Rubin Telescope.    

\begin{figure}
    \centering
    \includegraphics[width=0.98\linewidth]{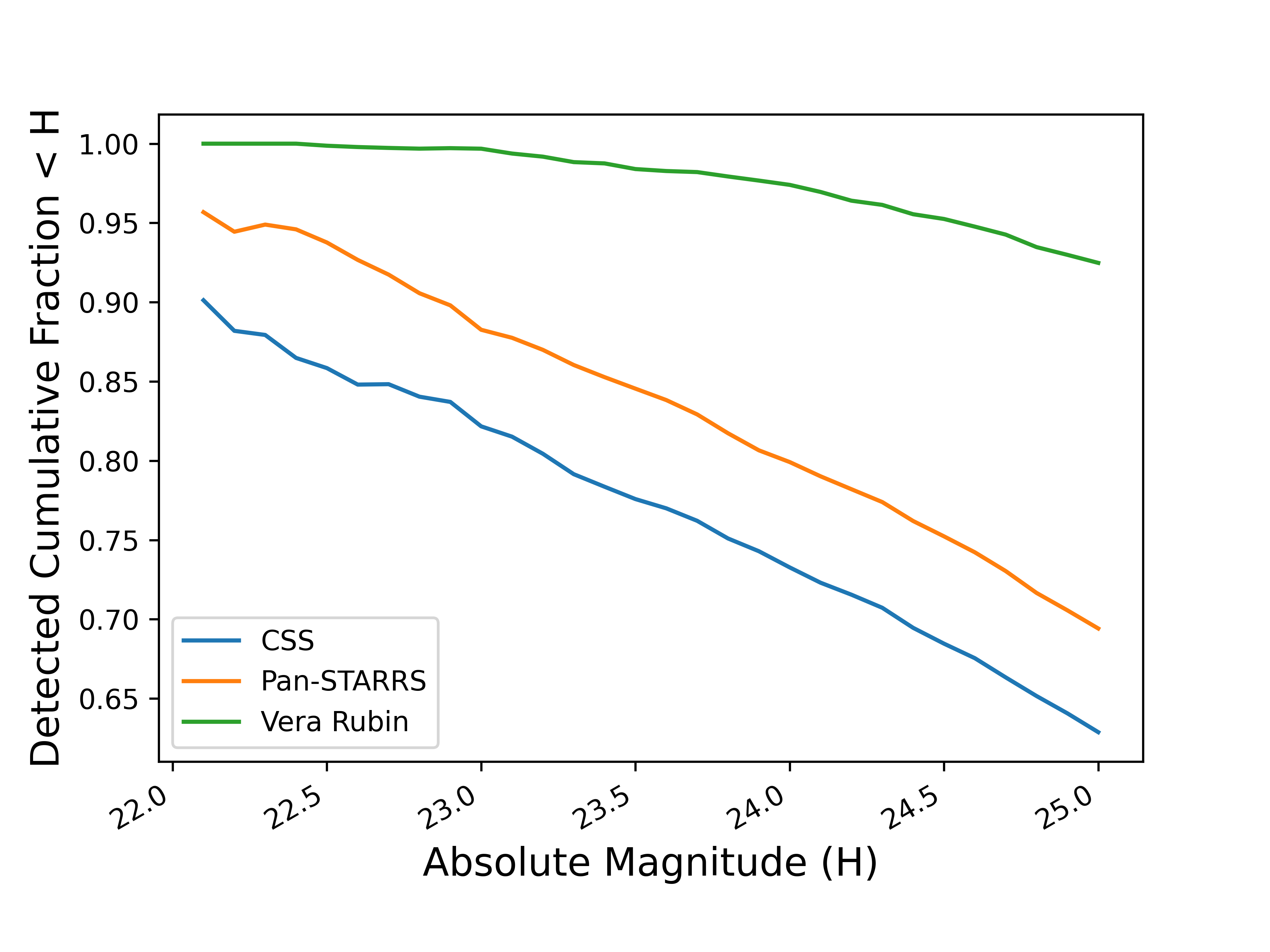}
    \caption{Detected cumulative fraction of Kamo`oalewa-like objects obtained at the end of the simulations, as a function of the absolute magnitude $H$.}
    \label{fig:survey_other}
\end{figure}

The left panel of Fig~\ref{fig:survey_other} shows the cumulative fraction of discovered quasi-satellites at the end of the simulation timespan as a function of the absolute magnitude $H$. In the interval $22< H <25$, the fraction of discovered objects shows a roughly linear decreasing trend, showing that smaller Earth quasi-satellites are harder to discover. The completeness of CSS at magnitude 22 reaches about 90\%, while it drops to 62\% at magnitude 25. Pan-STARRS is more efficient at smaller diameter, reaching about 70\% completeness at magnitude 25. Note that, at the same magnitude of Kamo`oalewa, Pan-STARRS has a discovery efficiency of about 80\%, meaning that it would be able to find an extremely large fraction of Kamo`oalewa-like objects. On the other hand, Vera Rubin has an efficiency of 90\% or larger through the whole absolute magnitude interval, showing its extreme efficiency in finding Earth quasi-satellites. 

\section{Discussion}
\label{s:discussion}

\subsection{Implications on the origin of Kamo`oalewa}
\begin{figure}
    \centering
    \includegraphics[width=0.48\textwidth]{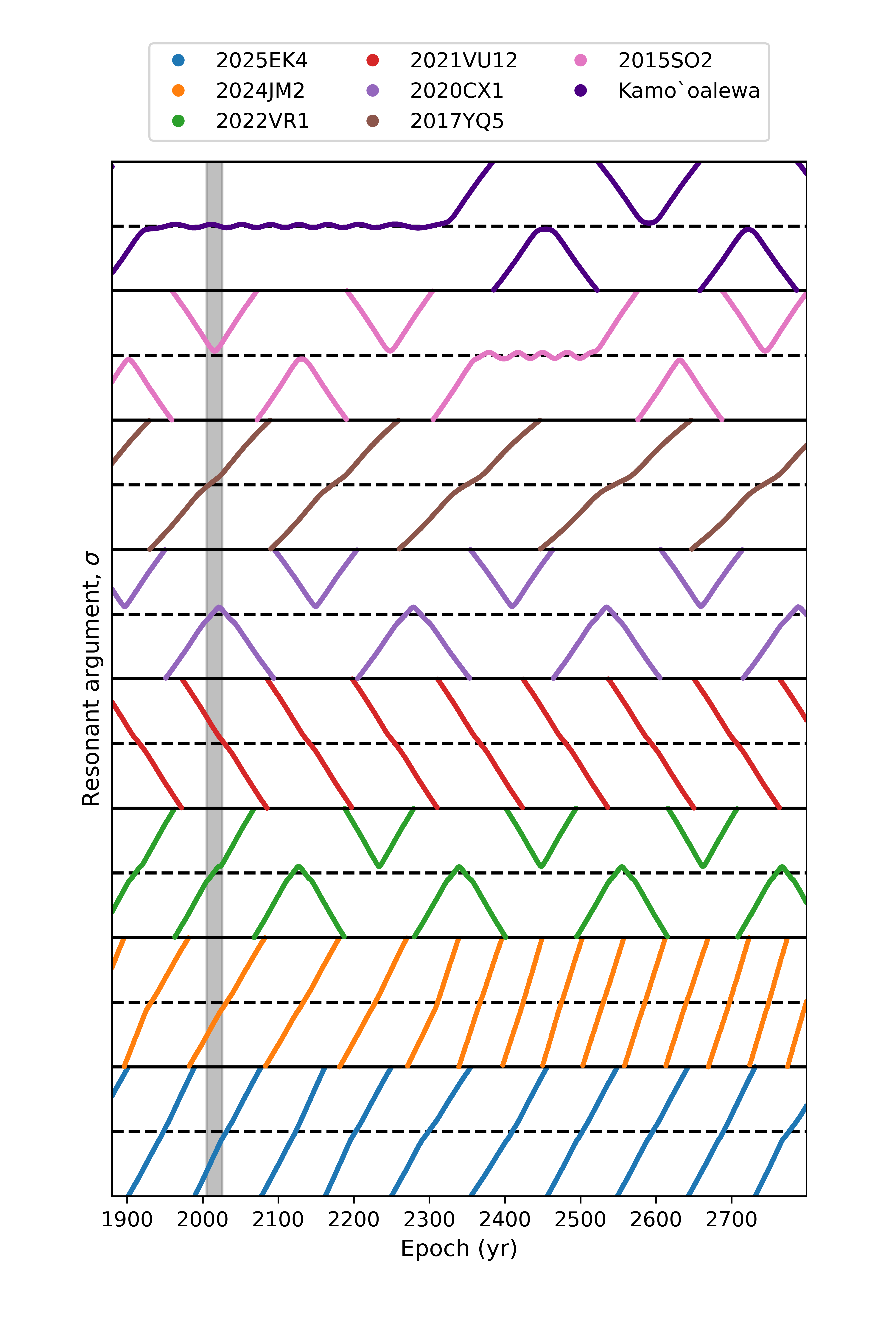}
    \caption{Evolution of the resonant argument $\sigma$ of the 8 known Earth co-orbitals with $22<H<25$, between 1880 and 2880. Each panel is an angle between $-180$ and 180 deg. The dashed black line corresponds to $\sigma = 0$ deg in each panel. The grey shaded area highlights the period from 1 January 2005 to 1 January 2026.}
    \label{fig:known_coorbitals}
\end{figure}

The estimates presented above show that, on average, we expect $1.23 \pm 0.13$ Kamo`oalewa-like objects at any time. At magnitude 24, the expected discovery efficiency of Earth quasi-satellites for Pan-STARRS is of 80\%, meaning that we expect $0.80 \times 1.23 \sim 1$ of such known objects. 
To understand how this estimate fits with the known data, we start from the known Earth co-orbitals and extract the asteroid which are currently quasi-satellites.To this end, we queried the NEA database of the ESA NEO Coordination Centre\footnote{\url{https://neo.ssa.esa.int/}} (NEOCC) to get all the known NEAs with absolute magnitude $H$ between 22 and 25, and fulfilling the constraints of Eq.~\eqref{eq:coorbital_space}. The database was queried on 21 November 2025. A total of 8 Earth co-orbitals were obtained (see Tab.~\ref{tab:neo_coorbitals}), including Kamo`oalewa. Their orbits were determined with the Aegis system \citep{fenucci-etal_2024b} and, to check their resonant behaviour, we propagated their orbits with the high fidelity dynamical model for Solar System small bodies implemented in Aegis.
Figure~\ref{fig:known_coorbitals} shows the evolution of the critical argument $\sigma$, between year 1880 and 2880. Four of these objects are not in resonance with the Earth, and 3 of them are currently on a horseshoe orbit, leaving Kamo'oalewa as the only quasi-satellite currently known in its size range. Note that 2015~SO$_2$ experiences a quasi-satellite configuration longer than 100 years, but starting around year 2350 \citep[see also][]{delafuente-delafuente_2016b}, and that Kamo`oalewa will switch to horseshoe state shortly before that epoch. This gives credibility to the models' predictions that there should exist only one Kamo`oalewa-like object at any given time. The figure also highlights the period from 2005 to 2026 in grey, which roughly corresponds to the timespan of activity of CSS and Pan-STARRS. Note that all the objects have been discovered near their passage close to $\sigma = 0$ deg, which corresponds to a small difference in mean longitude of the asteroid and the Earth. Thus, the estimates presented here suggest that the current population of Earth quasi-satellites with the same size and orbital properties of Kamo`oalewa could be already complete, and it can be explained without relying on the Lunar origin hypothesis.

Going down to absolute magnitude 26, there are two others known Earth quasi-satellites: 2025~PN$_7$ and 2023~FW$_{13}$ \citep{delafuente-delafuente_2025}. The linear decreasing trend of the survey efficiency shown in Fig.~\ref{fig:survey_other} cannot be extrapolated to fainter magnitudes, because at some point there would necessarily be an exponential decay. For this reason, we took all the orbits generated to obtain the results of Sec.~\ref{ss:completion_results} and assigned them an absolute magnitude of 26, and re-run the survey simulations for Pan-STARRS. This way, we obtained a completeness limit of 14\% reached in 2022. On the other hand, NEOMOD3 predicts a $2427 \pm 50$ Earth co-orbitals with $H<26$ and, since the estimate of Sec.~\ref{s:results} do not depend on the size, this converts to about 30 expected Earth quasi-satellites. Combining this number with the discovery efficiency of Pan-STARRS, we would expect $30 \times 0.14 \sim 4$ known Earth quasi-satellites with $H<26$. This is roughly in agreement with the known population, which can again be explained without relying on Lunar ejecta. 
 
The results obtained for the Vera Rubin Observatory show that this survey will provide data to assess the validity of our estimates. On the other hand, the general population of Earth co-orbitals is more difficult to discover from ground-based observatories only \citep[see also][]{morais-morbidelli_2002}, because the $\sigma$ period of horseshoe and circulating orbits is typically of the order of hundreds of years. The upcoming space-based infrared telescopes NEO Surveyor \citep{mainzer-etal_2023} by NASA and NEOMIR by ESA \citep{conversi-etal_2023} may be able to discover a larger part of the remaining Earth co-orbital population, since they are designed to observe at lower solar elongations.  

{\renewcommand{\arraystretch}{1.2}
\begin{table}[ht!]
\centering
\caption{Known NEAs in the Earth co-orbital space with $22<H<25$ extracted from the NEOCC database.}
\begin{tabular}{lcccc}
\hline
\hline
Object & $a$ (au) & $e$ & $i$ (deg) & $H$ \\
\hline
2015~SO$_2$ & 0.994223 & 0.10880   & 9.1672  & 23.699 \\
2017~YQ$_5$ & 0.995835 & 0.168818 & 16.9249 & 24.725 \\
2020~CX$_1$ & 1.004272 & 0.163259 & 12.7317 & 24.080 \\
2021~VU$_{12}$ & 1.004866 & 0.147931 & 17.5463 & 24.193 \\
2022~VR$_1$ & 0.994258 & 0.167639 & 5.8289  & 24.738 \\
2024~JM$_2$ & 0.994981 & 0.199867 & 19.9947 & 23.827 \\
2025~EK$_4$ & 0.994482 & 0.193721 & 17.6789 & 24.014 \\
Kamo`oalewa  & 1.000905 & 0.102447 & 7.8000  & 24.008 \\
\hline
\end{tabular}
\label{tab:neo_coorbitals}
\end{table}
}

In contrast, the optimistic estimates presented in Sec~\ref{s:results} yield an optimistic number of $0.042$ Kamo`oalewa-like objects originating as Giordano Bruno ejecta, which becomes even slightly lower if we take into account the efficiency of the currently operating surveys. 
Thus, these numbers suggest that a main-belt origin is statistically more likely than the Giordano Bruno origin hypothesis, and this is consistent with the long space-weathering exposure scenario proposed by \citet{zhang-etal_2024}, which therefore appears plausible.
Initial in-situ observations from the Tianwen-2 spacecrafts should be able to give an better assessment of the origin of Kamo`oalewa, and we anticipate that a definitive answer will be provided by laboratory analyses of the samples returned to Earth. The returned samples would allow a better and deep comparison with the Luna 24 samples containing ejected material from the Giordano Bruno crater \citep{basilevsky-head_2012}, allowing a direct testing of this particular hypothesis on the origin. 

If Kamo`oalewa is actually a Lunar ejecta as proposed by \citet{sharkey-etal_2021,jiao-etal_2024}, it would be the first ever detected NEA to originate from the Moon, marking an extraordinary low probability discovery which would open new paradigms for NEA formation and population models (see Sec.~\ref{ss:limitations}). 
On the other hand, if Kamo`oalewa originated in the main belt as it was shown here to be more likely, then the next challenges would be to reconstruct its past evolution, identify possible parent bodies in the main belt, and eventually corroborating the LL-chondrite analogue through laboratory analyses of the returned samples. This possibility would also provide a better understanding of the formation process and evolution of small asteroids, which are still poorly constrained, giving a more comprehensive understanding of our Solar System.  

\subsection{Lunar Ejecta in the context of NEA population models}
\label{ss:limitations}
Current NEA population models assume that the main belt is the principal source of NEAs. Other low-order effects are also included, such as the catastrophic disruption at low-perihelion distance \citep{granvik-etal_2016}, and the tidal disruption at close approaches with terrestrial planets \citep{granvik-walsh_2024, nesvorny-etal_2024}. Note however that the tidal disruption has a significant effect only at $H > 25$, thus at sizes smaller than that of Kamo`oalewa. On the other hand, lunar ejecta are not accounted for as a source of NEAs. 
It is worth noting that the calibration of the model by \citet{granvik-etal_2018} was performed with NEA data obtained between 2005 and 2012, all from the CSS survey. In this timespan the known NEA population comprised less than 10~000 NEAs, thus much more limited than that of today. The only asteroids proposed to be Lunar ejecta so far (i.e. Kamo`oalewa, 2020~CD$_3$ \citet{bolin-etal_2020}, and 2024~PT$_5$ \citet{kareta-etal_2025}) were not discovered yet at the time the model was developed. Moreover,  \citet{wu-etal_2025} recently estimated the expected annual number of detections of NEAs of Lunar origin. The authors obtained 0.75 annual discoveries for Pan-STARRS and 0.31 for the Asteroid Terrestrial-impact Last Alert System (ATLAS). While CSS was not taken into account in their study, it is reasonable to assume a number between 0.31 and 0.75 for CSS. This  means that the dataset used to calibrate the model by \citet{granvik-etal_2018} would possibly contain $\sim 5$ NEAs of Lunar origin, which is probably too few to have introduced any significant bias in the calibration of the NEA population model.  The agreement with the NEOMOD3 model calibrated with a larger and more recent sample of data is another hint that Lunar ejecta should not constitute a significant fraction of the NEA population, at least at decametres and larger scale.  

Recent estimates by \citet{wu-etal_2025} shows that there could be about 500~000 NEAs larger than 5 m originating from the Moon as impact ejecta, which is still less than 1\% of the total expected NEA population. 
The recently discovered asteroid 2024~YR$_4$ \citep{2024MPEC....Y..140D} currently has a probability of 4.3\% of impacting the Moon in 2032 \citep{farnocchia-etal_2025}, highlighting that these stochastic events are a current concern \citep{wiegert-etal_2025}. Consequently, although the contribution of Lunar ejecta is almost negligible, modelling Lunar impact ejecta and incorporating them in NEA population models could be a useful exercise in the next future.  

Developments in this direction were recently presented by \citet{jedicke-etal_2025}, who put constraints on the distribution of Earth's minimoons and temporarily captured objects of Lunar origin. Although their estimates were affected by significant uncertainties on the frequency of impacts and their parameters, the authors found that Lunar ejecta could give rise to 1-6 temporarily captured object with a size of $\sim$1 m per year. They also found that this distribution quickly drops as the diameter grows, and the most optimistic estimate for a 20 meter temporarily captured object with Lunar origin is 1 every 1000 years. While this study was focused on Earth's minimoons and not on the whole Earth co-orbital region, it already shows the quick depletion of the ejecta population at decametres size.    
For larger asteroids, i.e. larger than $\sim$20 m in diameter, the contribution of Lunar ejecta to the whole NEA population has not been clearly quantified. The ejection of fragments of this size typically requires an impactor larger than about 500 meters \citep{singer-etal_2020}. The impact frequency of such objects from the NEA models by \citet{granvik-etal_2018, nesvorny-etal_2024b} is $\sim$6 Myr$^{-1}$ for the Earth. By scaling with the Lunar cross section, the impact frequency for the Moon is $\sim$0.45 Myr$^{-1}$, which is comparable to the average lifetime of NEAs. 

\section{Conclusions}
\label{s:conclusions}
Our study provides a quantitative assessment of the potential origin of asteroid (469219) Kamo`oalewa through dynamical simulations and current NEA population models. By evaluating the fraction of Earth's quasi-satellites within the Earth's co-orbital region and applying this to main belt-derived NEA population models, we estimate that the number of Kamo`oalewa-like objects originating from the main belt is $1.23 \pm 0.13$, and it is more than one order of magnitude larger than those potentially originating as Giordano Bruno ejecta. 
We also analysed the discovery efficiency of Kamo`oalewa-like objects of CSS, Pan-STARRS, and the Vera Rubin Observatory. We found a discovery efficiency of between 95\% and 70\% for absolute magnitude between 22 and 25 for Pan-STARRS, which is in agreement with the known population of Earth quasi-satellites. The Vera Rubin Observatory should reach a completeness of 95\% down to absolute magnitude 25, while CSS shows lower discovery efficiency, but still larger than 60\%.

These findings suggest that Kamo`oalewa is more likely to have originated from the main asteroid belt, and that it reached later the Earth co-orbital region. This interpretation aligns with the recent spectral analyses of laser irradiated LL-chondrites showing a steep red slope and an absorption band at around 1 $\mu$m, both compatible with the data about the spectrum of Kamo`oalewa.  
While the lunar origin hypothesis cannot be entirely ruled out, especially given Kamo`oalewa’s unique spectral features, our results highlight the importance of a probabilistic and dynamical framework in assessing the origin of NEAs. 
Observations by JWST in the first quarter of 2026 are going to provide additional spectral data, and upcoming in-situ observations and eventual sample returned by the Tianwen-2 mission are expected to provide definitive constraints on Kamo`oalewa’s composition and origin, offering a crucial test for the hypotheses presented here.

\begin{acknowledgements}
M. F. dedicates this work to the memory of Professor Àngel Jorba. This research has made use of data and/or services provided by the International Astronomical Union's Minor Planet Center. P. Z. was supported by the National Natural Science Foundation of China, Grant U24A2008.
\end{acknowledgements}
\bibliographystyle{aa}
\bibliography{holybib.bib}{}

\end{document}